\documentclass[aps,prl,twocolumn]{revtex4}
\usepackage{amsfonts}
\usepackage{amsmath}
\usepackage{graphicx}
\usepackage{dsfont}
\usepackage{diagbox}
\usepackage{array}
\usepackage{amssymb}
\usepackage{bbm}
\usepackage{float}
\usepackage{subfigure}
\usepackage{url}
\usepackage{hyperref}
\usepackage{bm}

\newcommand{\PreserveBackslash}[1]{\let\temp=\\#1\let\\=\temp}
\newcolumntype{C}[1]{>{\PreserveBackslash\centering}p{#1}}
\newcolumntype{R}[1]{>{\PreserveBackslash\raggedleft}p{#1}}
\newcolumntype{L}[1]{>{\PreserveBackslash\raggedright}p{#1}}

\begin{document}

\title{Tensor network approach to phase transitions of a non-Abelian
topological phase}
\author{Wen-Tao Xu$^{1}$, Qi Zhang$^{1}$ and Guang-Ming Zhang$^{1,2}$}
\affiliation{$^{1}$State Key Laboratory of Low-Dimensional Quantum Physics and Department
of Physics, Tsinghua University, Beijing 100084, China\\
$^{2}$Frontier Science Center for Quantum Information, Beijing 100084, China.}
\date{\today}

\begin{abstract}
The non-abelian topological phase with Fibonacci anyons minimally supports
universal quantum computation. In order to investigate the possible phase
transitions out of the Fibonacci topological phase, we propose a generic
quantum-net wavefunction with two tuning parameters dual with each other,
and the norm can be exactly mapped into a partition function of the
two-coupled $\phi^{2}$-state Potts models, where $\phi =(\sqrt{5}+1)/2$ is
the golden ratio. By developing the tensor network representation of this
wavefunction on a square lattice, we can accurately calculate the full
phase diagram with the numerical methods of tensor networks. More
importantly, it is found that the non-abelian Fibonacci topological phase is
enclosed by three distinct non-topological phases and their dual phases of a
single $\phi^{2}$-state Potts model: the gapped dilute net phase, critical
dense net phase, and spontaneous translation symmetry breaking gapped phase.
We also determine the critical properties of the phase transitions among the
Fibonacci topological phase and those non-topological phases.
\end{abstract}

\maketitle

\textit{Introduction}. - In recent years theoretical and experimental search
for topological quantum phases of matter with anyonic excitations have
attracted considerable attention, because non-abelian quasiparticles are
necessary ingredient for topological quantum computation\cite%
{kitaevToricCode2003,freedman2003,kitaevHoneycomb,RMP_TQC,Weneaal3099}.
Since the non-abelian topological phases are characterized by fractionalized
degrees of freedom\cite{kitaevHoneycomb,Chen_Gu_Wen_2010}, the
Landau-Ginzburg-Wilson theory cannot be used to characterize these exotic
phases, and their phase transition to other non-topological phases is an
important open problem.

One remarkable feature of topological phases is that the ground-state wave
function encodes many of the quasiparticle properties, which was exploited
as far back as the Laughlin's pioneering work on fractional quantum Hall
effect\cite{LaughlinWaveFunction}. Many properties of the topological phases
can also be deduced by mapping the wavefunction to a statistical mechanics
model. In this paper, we will develop a tensor network approach by
constructing a generic topological wavefunction with tuning parameters,
which directly encodes the topological properties in the virtual symmetries
of the local tensor. By studying the corresponding partition function, we can
detect possible topological phase transitions and identify the associated
anyon-condensation mechanism.

The Fibonacci anyon phase is the simplest one supporting universal quantum
computation. The Fibonacci anyon $\tau $ obeys the non-abelian fusion rule:
$\tau \otimes \tau =1\oplus \tau $, where $1$ is the trivial particle, and one
Fibonacci anyon carries a non-integer quantum dimension $\phi =(\sqrt{5}+1)/2$.
A prototype lattice model realizing the Fibonacci anyons is the Levin-Wen
string-net model\cite{levin_string-net_2005} with additional two types of
anyons: $\bar{\tau}$ with the opposite chirality to $\tau $ and a bosonic
composite particle $b=\tau \otimes \bar{\tau}$. This string-net model just
represents the fixed point of the doubled Fibonacci (DFib) topological phase
with zero correlation length. To consider the topological phase transitions
out of the DFib topological phase, one has to drive the string-net model
away from its fixed point by introducing a string tension\cite%
{J_Vidal_Golden_string_net,Vidal2015,schotte_tensor-network_2019,Condensation_driven}.
However, due to the lack of quantum self-duality, a generic phase diagram of
the DFib topological phase has not been obtained.

\begin{figure}[tbp]
\centering
\includegraphics[width=7cm,]{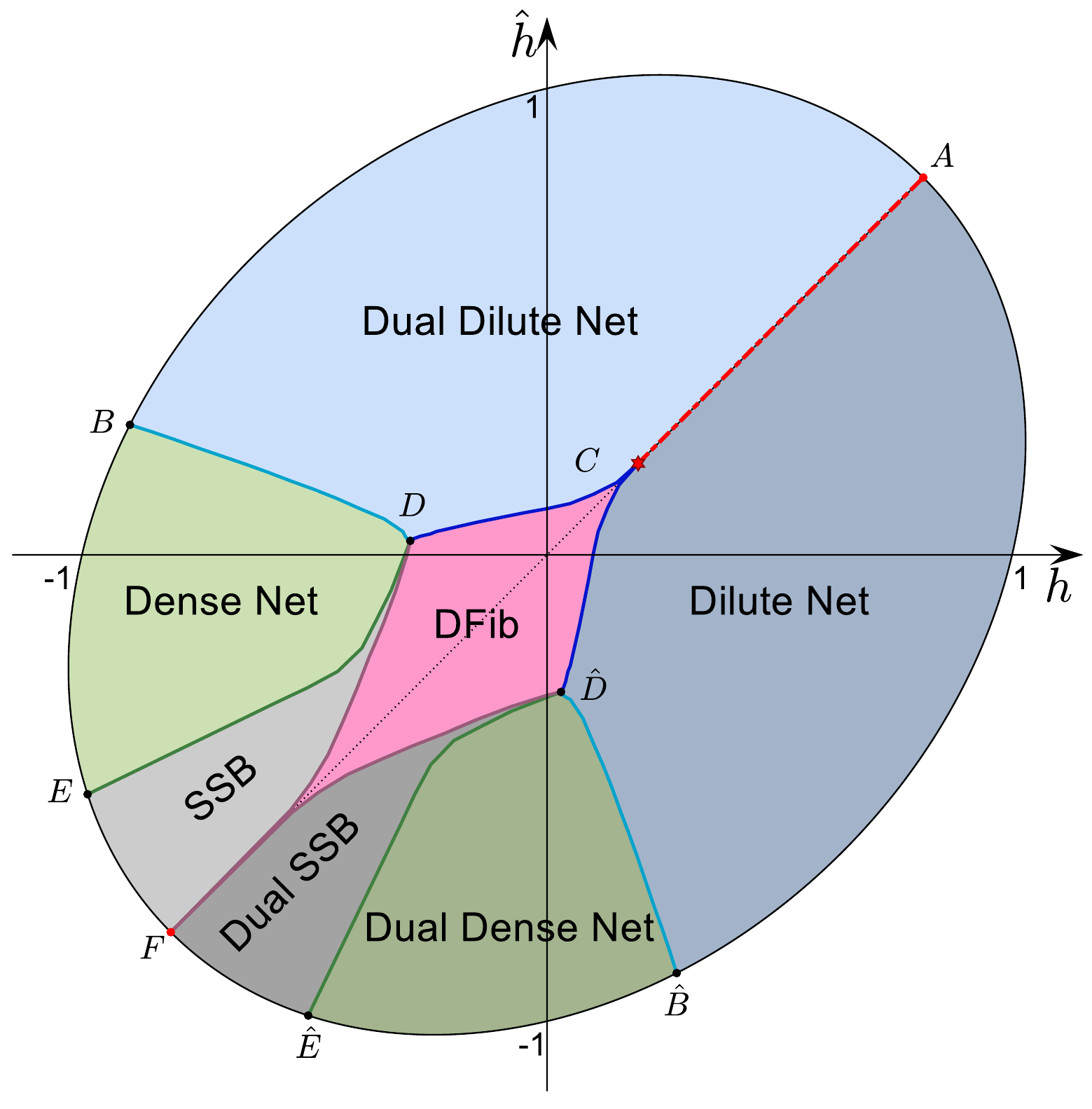}
\caption{The global phase diagram for the generic quantum-net phase with two
dual parameters. The phase diagram is symmetric about the self-dual line $h=%
\hat{h}$ and bounded by the elliptic curve of the two-decoupled Potts
models. The $AC$ dot-dashed line represents a weak first-order transition
and all other solid lines are continuous transitions. $C$ denotes a
tri-critical point and $D$ is a tetra-critical point. }
\label{phase_diagram}
\end{figure}

It was noticed that a quantum-net model is suggested to describe the DFib
topological order with a finite correlation length on a square lattice\cite%
{fendley_topological_2008,fendley_fibonacci_2013}. The most important
feature is the presence of quantum self-duality. In this paper, we propose a
generic DFib quantum-net wavefunction with \emph{two} dual string tensions%
\cite{xu2019tricritical}, and its norm can be mapped into a partition
function of the two-coupled $\phi^{2} $-state Potts models, whose Boltzmann
weights can be negative. In order to study the quantum topological phase
transitions numerically, we derive the triple-line tensor network state
(TNS) representation of this generic wavefunction. Then the global phase
diagram is fully established using the corner transfer matrix (CTM) method%
\cite{CTM_Nishino,CTM_Vidal,CTM_Corboz,fishman_faster_2018} and variational
uniform matrix product state (VUMPS) method\cite%
{Tangent_space,fishman_faster_2018,zauner-stauber_variational_2018}. As
shown in Fig. \ref{phase_diagram}, the non-abelian DFib phase is present
only in the two-coupled Potts models and surrounded by three distinct
non-topological phases and their dual phases: the gapped dilute net phase,
critical dense net phase, and spontaneous symmetry breaking (SSB) gapped
phase.

\begin{figure}[tbp]
\centering
\includegraphics[width=8cm,trim=40 65 0 65,clip]{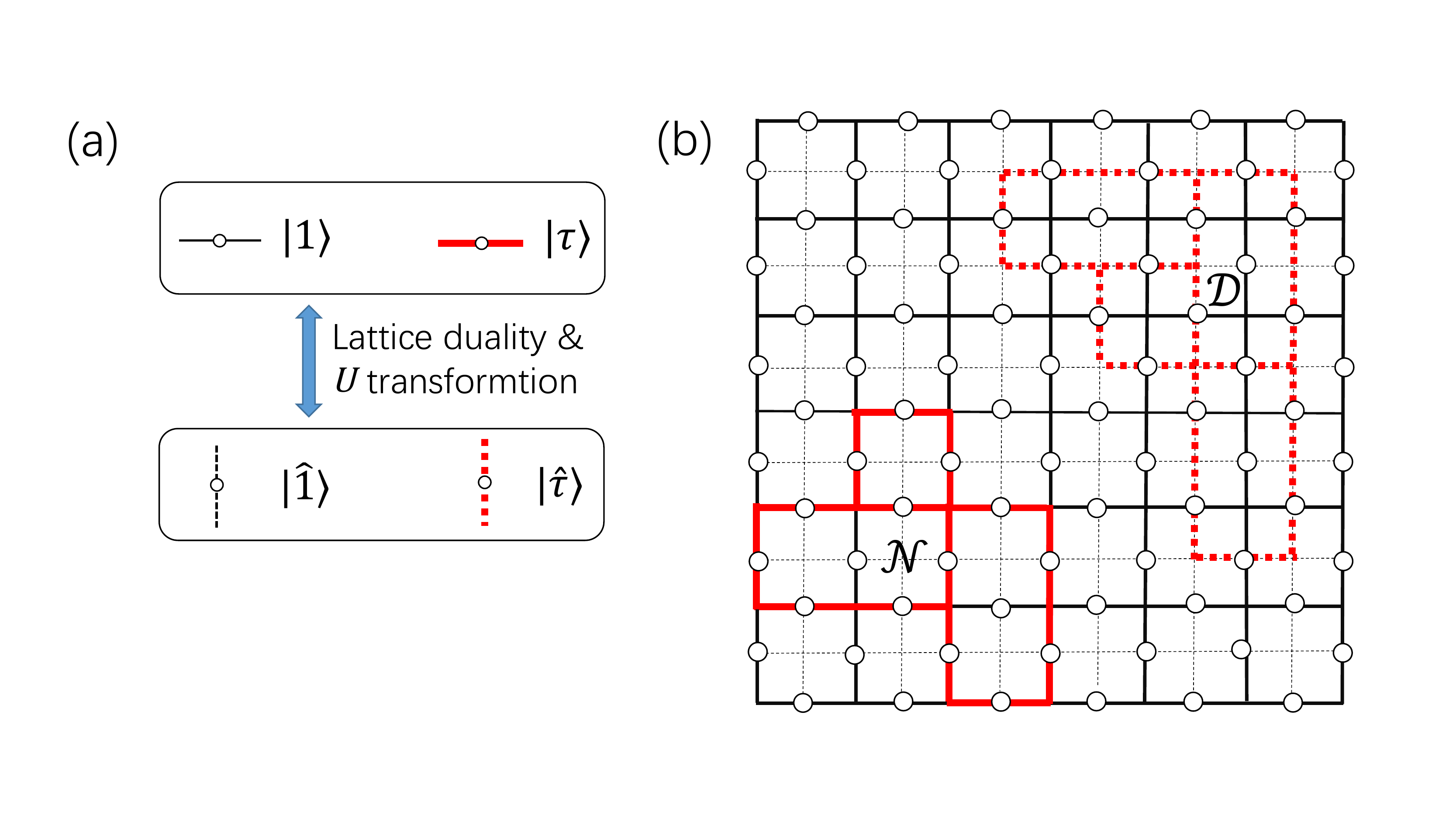}
\caption{(a) Two local orthogonal quantum states $|1\rangle $ and $|\protect%
\tau \rangle $ and their dual local states $|\hat{1}\rangle $ and $|\hat{%
\protect\tau}\rangle $. (b) The physical degrees of freedom locate at the
edges of the square lattice. The full lines denote the original lattice
while the dash lines are dual lattice. $\mathcal{N}$ and $\mathcal{D}$ are
two typical nets. }
\label{Net}
\end{figure}

\textit{Generic quantum-net wavefunction}. - The quantum-net wavefunction
involves nets and chromatic polynomials\cite{fendley_fibonacci_2013}. An
edge of the lattice is either empty or occupied by $\tau $ string, yielding
two mutually orthogonal quantum states $|1\rangle $ and $|\tau \rangle $,
see Fig. \ref{Net} (a). A $\tau$ string consists of the $|\tau\rangle$
states on the connected edges, and a net $\mathcal{N}$ is formed by the
closed $\tau $ strings which are allowed to branch and cross, as shown in
Fig. \ref{Net} (b). Since the net $\mathcal{N}$ divides the two-dimensional
manifold into different regions, the chromatic polynomial $\chi _{\hat{%
\mathcal{N}}}(Q)$ with $Q\in\mathbb{N}_{+}$ counts the ways of coloring the
net $\mathcal{N}$ using $Q$ different colors (nets $\mathcal{N}$ and $\hat{%
\mathcal{N}}$ are dual each other), such that the neighboring regions
sharing a boundary are colored differently. Since $\chi _{\hat{\mathcal{N}}%
}(Q)$ is a polynomial of $Q$, it can be generalized to $Q\in\mathbb{R}$.

On a square lattice, the DFib quantum-net wave function\cite%
{fendley_topological_2008,fendley_fibonacci_2013} is given by the
superposition of nets $\mathcal{N}$:
\begin{equation}
|\Psi \rangle =\sum_{\mathcal{N}}\phi ^{-L_{\mathcal{N}}/2}\chi _{\hat{%
\mathcal{N}}}\left( \phi ^{2}\right) |\mathcal{N}\rangle ,
\label{Fendley_wavefunction}
\end{equation}%
where $L_{\mathcal{N}}$ is the total length of the $\tau $ strings in the
net $\mathcal{N}$ and $\phi^{-L_{\mathcal{N}}/2}$ is viewed as a string
tension. The end of $\tau$ string will carry a Fibonacci anyon with
fractionalized quantum dimension $\phi$. To show the quantum self-duality,
the wave function $|\Psi\rangle$ has to be written on the dual lattice in
terms of the orthogonal quantum states $|\hat{1}\rangle $ and $|\hat{\tau}%
\rangle $, which are related to the local states $|1\rangle $ and $%
|\tau\rangle $ via the transformation\cite{fendley_topological_2008}:
\begin{equation}
U=\left(
\begin{array}{cc}
\langle1|\hat{1}\rangle & \langle\tau|\hat{1}\rangle \\
\langle1|\hat{\tau}\rangle & \langle\tau|\hat{\tau}\rangle%
\end{array}%
\right)=\frac{1}{\phi }\left(
\begin{array}{cc}
1 & \sqrt{\phi } \\
\sqrt{\phi } & -1%
\end{array}%
\right) .  \label{transformation}
\end{equation}
Then the dual quantum-net wavefunction has the same form as $|\Psi\rangle$.

Inspired by the deformed $\mathbb{Z}_2$ abelian topological phase\cite%
{GaugingTNS,zhu_gapless_2019}, we propose a generic quantum-net wavefunction
\begin{equation}
|\Psi (h,\hat{h})\rangle =\prod_{\text{edges}}P(h,\hat{h})|\Psi \rangle ,
\label{deformation}
\end{equation}%
where $P(h,\hat{h})=(1+h\sigma ^{z}+\hat{h}\hat{\sigma}^{z})$ is the
deformation matrix acting on all edges, $\sigma ^{z}$ is the diagonal Pauli
matrix in the $|1\rangle $ and $|\tau \rangle $ basis, and $\hat{\sigma}%
^{z}=U\sigma ^{z}U^{-1}$ is diagonal Pauli matrix in the $|\hat{1}\rangle $
and $|\hat{\tau}\rangle $ basis. $h$ and $\hat{h}$ describe the string
tensions. The quantum duality transforms $|\Psi (h,\hat{h})\rangle $ into $%
|\Psi (\hat{h},h)\rangle $, and the quantum self-duality exhibits when $h=%
\hat{h}$.

It should be emphasized that the generic wavefunction still has a local
parent Hamiltonian. As it is shown, the quantum-net $|\Psi \rangle $ has a
frustration-free parent Hamiltonian: $H=\sum_{a}H_{a}$ is a sum of local
positive projectors \cite{fendley_fibonacci_2013}. Since the deformation
matrix $P$ is a local positive definite operators, the parent Hamiltonian of
the generic wavefunction is given by $H(h,\hat{h})=%
\sum_{a}P_{a}^{-1}H_{a}P_{a}^{-1}$, where $P_a$ is a product of $P$ in the
support of $H_a$. The possible quantum critical points of this parent
Hamiltonian are characterized by the so-called conformal quantum critical
points\cite{ardonne,castelnovo}, where all equal-time correlators of local
operators are described by two-dimensional conformal field theories (CFTs).

\textit{Mapping to two-coupled Potts models}. - To extract the possible
quantum phase transitions out of the DFib topological phase, we consider the
norm of the generic quantum-net wavefunction,
\begin{equation}
\mathcal{Z}=\sum_{\mathcal{N},\mathcal{N}^{\prime }}\chi _{\hat{\mathcal{N}}%
}\left( \phi ^{2}\right) \chi _{\hat{\mathcal{N}}^{\prime }}\left( \phi
^{2}\right) \prod_{\text{edge}}W_{n,n^{\prime }},  \label{norm}
\end{equation}%
where $\mathcal{N}$ and $\mathcal{N}^{\prime }$ denote the nets in the bra
and ket layers and $n,n^{\prime }=1$ or $\tau $ corresponds to the empty or $%
\tau $-string occupied edge. The weight matrix $W$ is given by
\begin{equation}
W=W_{0}\left(
\begin{array}{cc}
1 & e^{-K} \\
e^{-K} & e^{-2K-K^{\prime }}%
\end{array}%
\right) ,  \label{Boltzmann_matrix}
\end{equation}%
where
\begin{eqnarray}
e^{-K} &=&\frac{4\hat{h}/\phi ^{2}}{4\hat{h}^{2}/\phi ^{3}+(1+h-\hat{h}/\phi
^{3})^{2}},  \notag \\
e^{-K^{\prime }} &=&\frac{4\hat{h}^{2}/\phi ^{3}+(1-h+\hat{h}/\phi ^{3})^{2}%
}{4\hat{h}^{2}/\phi ^{2}+\phi (1+h-\hat{h}/\phi ^{3})^{2}}e^{2K}.
\end{eqnarray}%
Since the $Q$-state Potts model is expressed as $\mathcal{Z}_{\text{Potts}%
}=\sum_{\mathcal{N}}e^{-\beta L_{\mathcal{N}}}\chi _{\hat{\mathcal{N}}}(Q)$,
$e^{-K}$ can be viewed as the Boltzmann weight for the $\phi ^{2}$-state
Potts model, while $K^{\prime }$ describes the inter-layer coupling.
Therefore, instead of the previous $(\phi+2)$-state Potts model\cite%
{fidkowski_string_2009}, the wavefunction norm (\ref{norm}) is mapped into
the partition function of two-coupled $\phi ^{2}$-state Potts models. Such a
statistical model is unusual in the sense that it involves negative
Boltzmann weight in the parameter region ($\hat{h}<0$). However, the
appearance of such minus forms pairs and will not spoil the solution to the
partition function.

Numerous exact results can be deduced. When the coupling of the two $\phi
^{2}$-state Potts models vanishes, the deformation matrix $P(h,\hat{h})$
just projects out the trivial product states\cite{zhu_gapless_2019}, i.e., $%
\det P(h,\hat{h})=0$, corresponding to an elliptic equation $h^{2}-2h\hat{h}%
/\phi ^{3}+\hat{h}^{2}=1$ as the boundary of the phase diagram (Fig. \ref%
{phase_diagram}). From this equation, we can find two self-dual points $h=%
\hat{h}=\pm \phi /2$. One is identified as the ferromagnetic critical point $%
A$ separating the gapped dilute net phase and its dual phase\cite%
{Saleur_1991}, and the other was considered as an "unphysical" point $F$ in
the previous study\cite{jacobsen_antiferromagnetic_2006}. However, we will
use the TNS methods to show that the point $F$ is a multi-critical point,
around which a new phase with translational symmetry breaking and its dual
phase are found\cite{SM}.

Along the self-dual line $h=\hat{h}$, it has been known that the transfer
operators of the two-coupled Potts models are endowed with $SO(4)_3$
Birman-Murakami-Wenzl algebra and a critical point $C\approx(0.197,0.197)$
has been found by the level-rank duality\cite{fendley_critical_2008}, which
is described by the coset CFT with a central charge $c=27/20$. As shown in
Fig. \ref{phase_diagram}, this critical point $C$ divides the self-dual line
into two parts: the DFib topological part $CF$ and the first-order phase
transition line $AC$ between the dilute net phase and its dual phase\cite{SM}%
.

\begin{figure}[tbp]
\includegraphics[width=8.5cm]{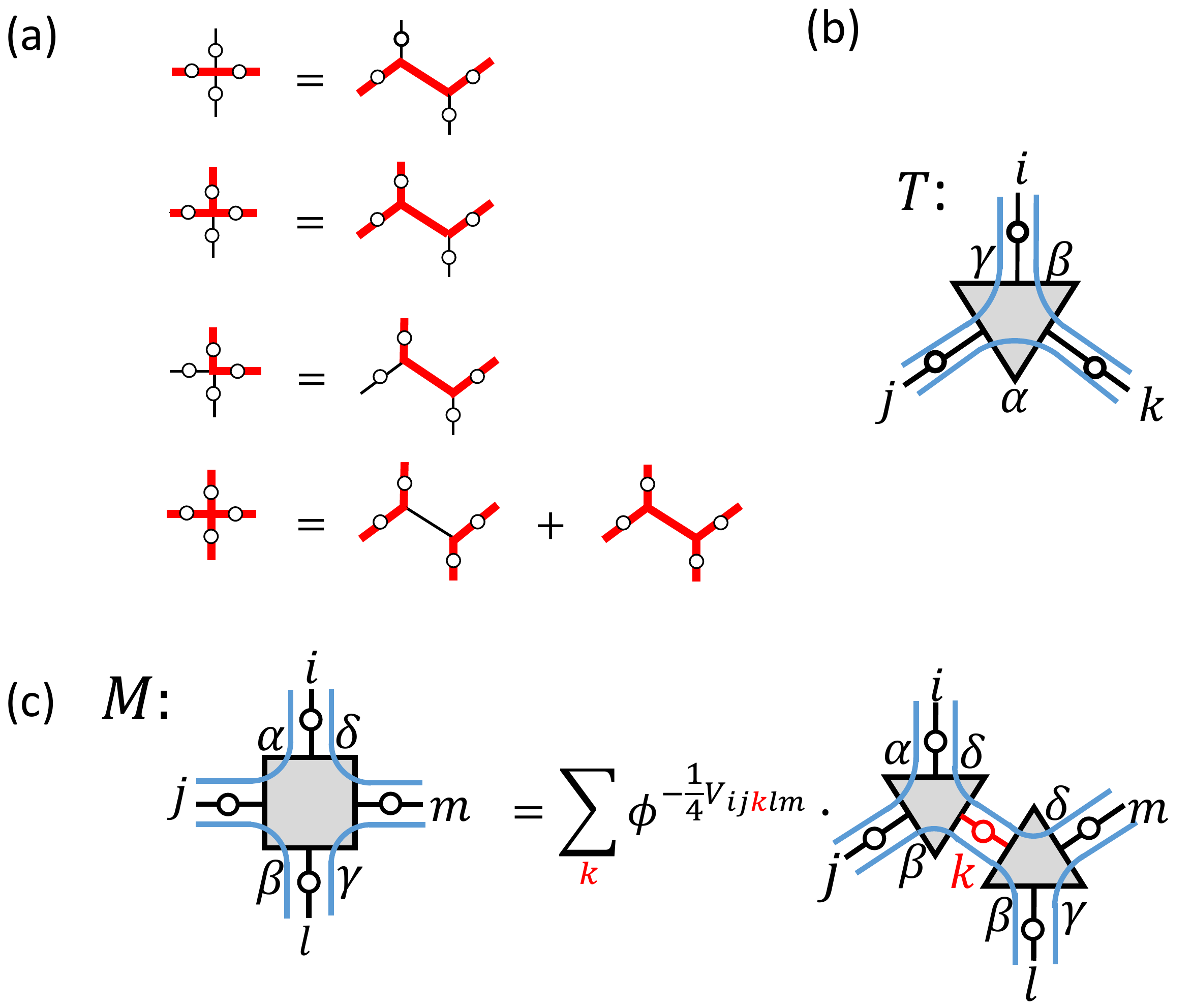}
\caption{(a) The square lattice is transformed to a honeycomb lattice. (b)
The triple line local tensor $T$ in the string-net wavefunction. (c)
Contracting two tensors $T$ yields the tensor $M $ on the square lattice.}
\label{squrae_tensor}
\end{figure}

\textit{Tensor network representation}. - In order to explore the large
parameter space of the global phase diagram, we have to employ the numerical
calculations. Before that, the TNS representation of the generic quantum-net
wavefunction should be established. Since the non-local chromatic
polynomials are involved in the wavefunction, auxiliary degrees of freedom
on the dual vertices are introduced to express the wavefunction in terms of
local structures. It is known that the TNS representation of the Levin-Wen
Fibonacci string-net $|\Psi _{\text{SN}}\rangle $ on the honeycomb lattice
has been established\cite{GuTensorNetwork,buerschaper_explicit_2009}. When
each vertex of a square lattice is split into two and added new edges with
no physical degrees of freedom, the square lattice is transformed into the
honeycomb lattice. Taking the advantage of the deletion-contraction relation%
\cite{fendley_topological_2008} displayed Fig. \ref{squrae_tensor} (a), we
can obtain the corresponding chromatic polynomial on the square lattice so
that the TNS for the quantum-net can be constructed.

As shown in Figs. \ref{squrae_tensor} (b) and (c), the local tensor of the
quantum-net on the square lattice is derived by contracting two triple-line
tensors $T_{\beta \delta \alpha }^{ijk}$ via
\begin{eqnarray*}
M_{\alpha \beta \gamma \delta }^{ijlm} &=&\sum_{k}\phi
^{-V_{ijklm}/4}T_{\beta \delta \alpha }^{ijk}T_{\delta \beta \gamma }^{lmk},
\\
V_{ijklm} &=&3\delta _{k\tau }\left( \delta _{i\tau }\delta _{j\tau }+\delta
_{l\tau }\delta _{m\tau }\right) +\delta _{i\tau }+\delta _{j\tau }+\delta
_{l\tau }+\delta _{m\tau }.
\end{eqnarray*}%
Then the TNS of the quantum-net wavefunction is expressed as
\begin{equation}
|\Psi \rangle =\sum_{\{\cdots ijlm\cdots \}}\text{tTr}\left[ \underset{\text{%
vertex}}{\bigotimes }M_{\alpha \beta \gamma \delta }^{ijlm}\right] |\cdots
ijlm\cdots \rangle ,
\end{equation}%
where \textquotedblleft tTr\textquotedblright\ denotes the tensor
contraction over all auxiliary indices. Finally, the deformed DFib
quantum-net wavefunction $|\Psi (h,\hat{h})\rangle $ is derived with the
modified local tensor $\widetilde{M}_{\alpha \beta \gamma \delta }^{ijlm} $,
which can be obtained by acting the deformation matrix $P$ on the physical
indices of $M_{\alpha \beta \gamma \delta }^{ijlm}$.

In the TNS representation, the topologically degenerated ground state can be
characterized with matrix product operators (MPOs) acting on the auxiliary
degrees of freedom\cite{AnyonMPO,PEPSTO,PEPS_as_ground_states}. The
quantum-net $|\Psi (h,\hat{h})\rangle $ shares the same MPOs as those of the
Fibonacci string-net, and the MPOs are independent of the deformation
parameters $h$ and $\hat{h}$. On a torus, the degenerate ground state space
is spanned by four minimal entangled states\cite{MES} labelled as $|\bm {1}%
\rangle $, $|\bm {\tau} \rangle $, $|\bm{\bar{\tau}}\rangle $ and $|\bm{b}%
\rangle $. With the MPO algebra\cite{AnyonMPO,Condensation_driven}, it can
be checked that our generic wavefunction $|\Psi (h,\hat{h})\rangle =|\bm{1}%
\rangle +|\bm{b}\rangle $. By inserting MPOs winding around the TNS
wavefunction $|\Psi (h,\hat{h})\rangle $, other three linear combinations of
the minimal entangled states can be obtained. In the generic TNSs, the
well-defined anyonic excitations can be created by simply manipulating on
the auxiliary degrees of freedom with MPOs, so one can easily measure the
condensation and confinement of anyons\cite{PEPSTO,Shadowofanyons}. To probe
the quantum phase transitions, it is sufficient to focus on $|\Psi(h,\hat{h}%
)\rangle $ without MPO insertion. However, if one concerns about the fates
of $\tau $ and $\bar{\tau}$ anyons, the TNSs with the MPO insertions\cite%
{TQFT-CFT} must be taken into consideration. Furthermore, the MPO symmetry
of the local tensor also constrains the possible CFTs describing conformal
critical points\cite{Buican2017,Anyon_chain}, and the CFTs must contain a
quantum dimension $\phi ^{2}$.

\textit{Global phase diagram}. - With the TNS for the deformed quantum net
at hand, those TNS algorithms can be applied. Integrating the physical
variables of the generic DFib wavefunction yields the partition function in
the form of a double-layer tensor network
\begin{equation}
\mathcal{Z}=\text{tTr}\left( \bigotimes_{\text{vertex}}\mathbb{M}\right) =%
\text{Tr}(\mathbb{T}^{L_{x}}),
\end{equation}%
where $\mathbb{M}$ is the local double triplet-line tensor obtained by
contracting the physical indices of $\widetilde{M}$ and its conjugate, $%
\mathbb{T}$ is the the column-to-column transfer operator, and $L_{x}$ is
the number of columns. In order to determine the various phase boundaries,
we need to calculate the correlation length, whose divergent peaks give rise
to the position of the continuous phase transitions. When the transfer
operator $\mathbb{T}$ is hermitian, we employ the VUMPS\cite%
{Tangent_space,fishman_faster_2018,zauner-stauber_variational_2018} to
extract the correlation length, while for non-hermitian $\mathbb{T}$ in the
lower half-plane of Fig.\ref{phase_diagram}, the CTM\cite%
{CTM_Nishino,CTM_Vidal,CTM_Corboz,fishman_faster_2018} method is used.

\begin{figure}[tbp]
\includegraphics[width=8.8cm]{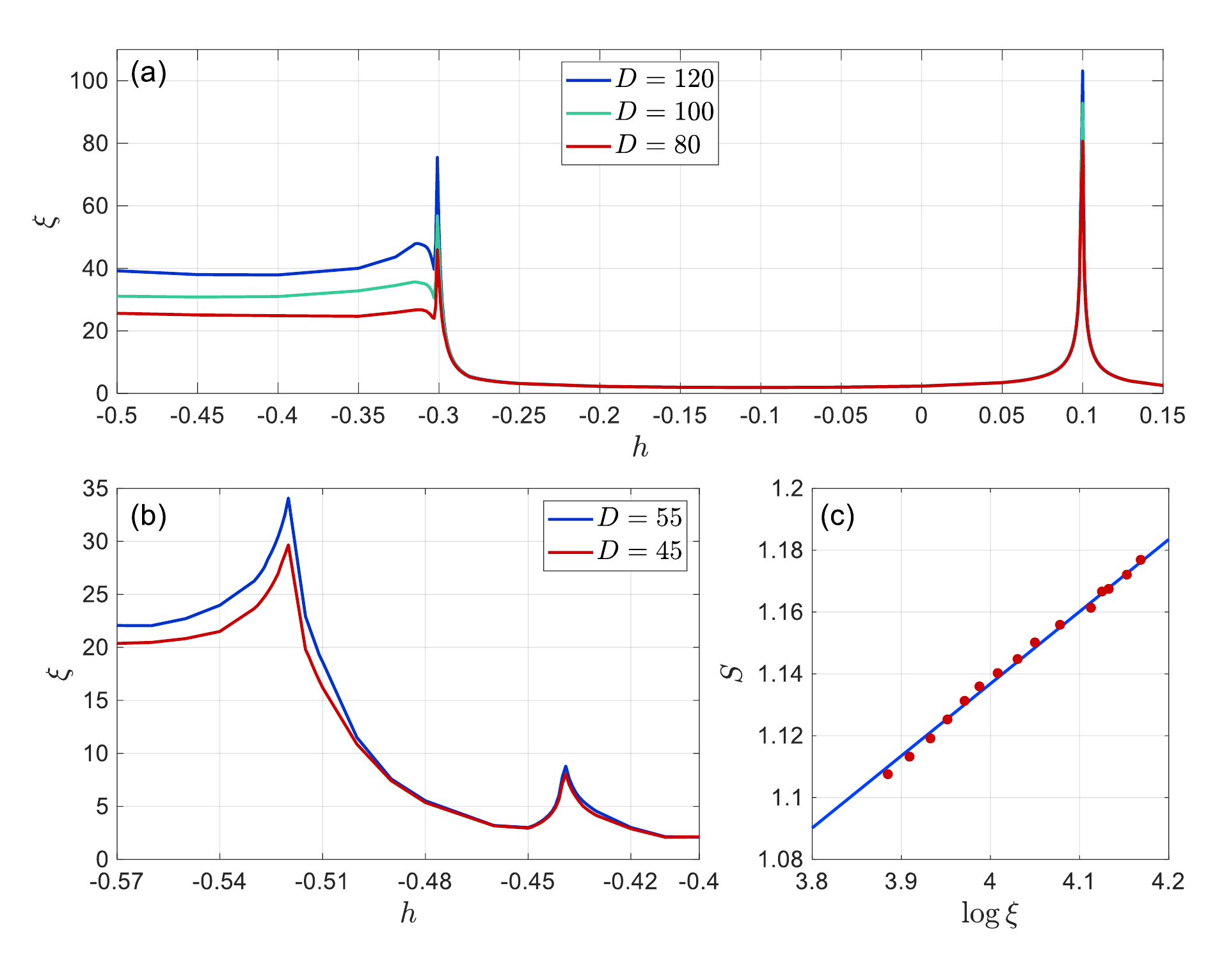}
\caption{(a) The correlation length $\protect\xi $ obtained by the VUMPS
method with bond dimensions $D$ along $h$ axis of the phase diagram. (b) The
correlation length deduced from the CTM method with bond dimensions $D$
along the cut $\hat{h}=-0.8-h$. $h=-0.4$ sits on the self-dual line. (c) The
entanglement entropy $S$ at the point $(h,\hat{h})=(-0.6,0)$. The blue line
is $S=\frac{c}{6}\log \protect\xi +S_{0}$ with the central charge $c=7/5$.}
\label{data}
\end{figure}

Along the axis of $h$, the numerical TNS calculation with VUMPS algorithm
has been performed with large bond dimensions $D=80,100,120$. As shown in
Fig. \ref{data} (a), there is a phase transition from the DFib topological
phase to the dilute net phase around $h\approx 0.1$. The peak position of
the correlation length is nearly the same for different bond dimensions. The
corresponding phase transition is still described by the CFT with a central
charge $c=14/15$, similar to that of the deformed DFib string-net\cite%
{Condensation_driven}. As $h$ is further increasing, the correlation length
first shows a divergent peak around $h\approx -0.3$ and then appears a hump,
which gradually becomes a broad peak with increasing the bond dimension.
After this hump, the system enters into a gapless dense net phase. The
finite entanglement scaling Fig. \ref{data} (c) suggests that the dense net
phase is described by the CFT with central charge $c=7/5$, corresponding to
the squared tri-critical Ising university class. Between the DFib
topological phase and the dense net phase, we find a narrow region of the
gapped phase with translational symmetry breaking. In order to clearly see
the SSB phase, we display the correlation length along the cut $\hat{h}%
=-0.8-h$ in Fig. \ref{data} (b), where the CTM methods are employed with the
bond dimensions $D=45,55$. The SSB phase exists between the two peaks of the
correlation length.

Moreover, the critical line $BD$ separating the dense net phase and dual
dilute net phase terminates at a special point $B=(-\sqrt{3-\phi }-1/\phi,%
\sqrt{3-\phi }-1/\phi )/2$, which is the critical point of the decoupled
antiferromagnetic $\phi ^{2}$-state Potts models characterizing by two
copies of $\mathbb{Z}_{3}$ parafermion CFT with $c=8/5$ (Ref. \cite%
{jacobsen_antiferromagnetic_2006}). But our numerics\cite{SM} shows that the
full line $BD $ is described by the CFT with the central charge $c\simeq 1.4$%
, as the same as the dense net phase. On the other hand, the critical line $%
ED$ separating the dense net phase from the SSB phase is described by the
CFT with the central charge $c\simeq 1.6$, while the continuous phase
transition line $FD$ between the DFib topological phase and the SSB phase is
characterized by the CFT with a central charge $c\approx 1.4$. Four critical
lines meet at a tetra-critical point $D\approx (-0.294,0.03)$. Using the
quantum duality, the complete phase diagram Fig. \ref{phase_diagram} is thus
fully established.

The universal feature of all gapped phases is the degeneracies of dominant
eigenvalues of the transfer operators with periodic boundary condition.
Since $|\Psi \rangle =|\bm{1}\rangle +|\bm{b}\rangle $, the $2$-fold
degenerate dominant eigenvalues of the DFib topological phase correspond to $%
\langle \bm{1}|\bm{1}\rangle $ and $\langle \bm{b}|\bm{b}\rangle $. Because $%
b$ anyons are condensed in the dilute net phase, the $4$-fold degenerate
eigenvectors belong to the topological sectors $\langle \bm{1}|\bm{1}\rangle
$, $\langle \bm{b}|\bm{b}\rangle $, $\langle \bm{1}|\bm{b}\rangle $ and $%
\langle \bm{b}|\bm{1}\rangle $. In the dense net phase, however, the large
nets dominate and $b$ anyons are logarithmically confined. Due to the
translation symmetry breaking in the SSB phase, there are two dominant
eigenvectors with the momenta $0$ and $\pi $ for each bra and ket of the
topological sectors, leading to $16$-fold degeneracy. All these results have
been confirmed in our numerical results\cite{SM}.

\textit{Conclusion}. - We have fully studied the non-abelian topological
phase transitions out of a generic DFib topological state. The norm of the
wave function is mapped into the partition function of the two-coupled $%
\phi^{2}$-state Potts models. With the tensor network representation and
numerical TNS methods, a global phase diagram has been fully established.
Previously we showed that\cite{zhu_gapless_2019} the $Z_2$ toric code
topological phase corresponds to the partial order phase of the
Ashkin-Teller model -- two-coupled Ising models. Here we further prove that
the non-abelian DFib topological phase can be mapped to an inter-layer
ordered phase of the two-coupled $\phi^{2}$-state Potts models, instead of
the the previous $(\phi+2)$-state Potts model\cite{fidkowski_string_2009}.

Compared to the Hamiltonian approach, such a wavefunction approach has many
advantages in applying the TNS methods to the quantum topological phase
transitions among intrinsic topological phases. A natural question is how
the conformal quantum criticality will be changed when the dynamics of the
parent Hamiltonian for the generic DFib topological phase is considered. The
related problems are under further investigations.

\textit{Acknowledgment}.- The authors would like to thank Hai-Jun Liao for
his providing the program of CTM algorithm and Guo-Yi Zhu for his
stimulating discussions. The research is supported by the National Key
Research and Development Program of MOST of China (2016YFYA0300300 and
2017YFA0302902).

\bibliography{ref}

\end{document}